\documentclass[aps,prx,twocolumn,superscriptaddress,10pt]{revtex4-2}

\usepackage{multirow,amsthm,amssymb,amsbsy,amsmath,epsfig,epstopdf,bm}

\usepackage{graphicx}
\usepackage{booktabs}
\usepackage{verbatim}
\usepackage{dcolumn}
\usepackage{color}
\usepackage[dvipsnames]{xcolor}
\usepackage[normalem]{ulem}
\usepackage{soul}
\newtheorem{theorem}{Theorem}

\newtheorem{lemma}[theorem]{Lemma}
\newtheorem*{theorem*}{Universality Theorem}

\newcommand{\RemoveSpaces}[1]{%
  \begingroup
  \spaceskip=1sp
  \xspaceskip=1sp
  #1%
  \endgroup}

\renewcommand{\onlinecite}[1]{\RemoveSpaces{[\citenum{#1}]}}

\setcitestyle{super} 

\begin{document}
\title{Gaussian states of continuous-variable quantum systems provide universal and versatile reservoir computing}

\author{Johannes Nokkala}
\email{johannes@ifisc.uib-csic.es }
\affiliation{Instituto de F\'{i}sica Interdisciplinar y Sistemas Complejos (IFISC, UIB-CSIC), Campus Universitat de les Illes Balears E-07122, Palma de Mallorca, Spain}

\author{Rodrigo Mart\'inez-Pe\~na}
\affiliation{Instituto de F\'{i}sica Interdisciplinar y Sistemas Complejos (IFISC, UIB-CSIC), Campus Universitat de les Illes Balears E-07122, Palma de Mallorca, Spain}

\author{Gian Luca Giorgi}
\affiliation{Instituto de F\'{i}sica Interdisciplinar y Sistemas Complejos (IFISC, UIB-CSIC), Campus Universitat de les Illes Balears E-07122, Palma de Mallorca, Spain}

\author{Valentina Parigi}
\affiliation{Laboratoire Kastler Brossel, Sorbonne Universit\'e, CNRS, ENS-PSL Research University, Coll\`ege de France; CC74 4 Place Jussieu, 75252 Paris, France}

\author{Miguel C. Soriano}
\affiliation{Instituto de F\'{i}sica Interdisciplinar y Sistemas Complejos (IFISC, UIB-CSIC), Campus Universitat de les Illes Balears E-07122, Palma de Mallorca, Spain}

\author{Roberta Zambrini}
\affiliation{Instituto de F\'{i}sica Interdisciplinar y Sistemas Complejos (IFISC, UIB-CSIC), Campus Universitat de les Illes Balears E-07122, Palma de Mallorca, Spain}

\date{\today}

\begin{abstract}		

We establish the potential of continuous-variable Gaussian states of linear dynamical systems for machine learning tasks. Specifically, we consider reservoir computing, an efficient framework for online time series processing. As a reservoir we consider a quantum harmonic network modeling e.g. linear quantum optical systems. We prove that unlike universal quantum computing, universal reservoir computing can be achieved without non-Gaussian resources. We find that encoding the input time series into Gaussian states is both a source and a means to tune the nonlinearity of the overall input-output map. We further show that the full potential of the proposed model can be reached by encoding to quantum fluctuations, such as squeezed vacuum, instead of classical intense fields or thermal fluctuations. Our results introduce a new research paradigm for reservoir computing harnessing the dynamics of a quantum system and the engineering of Gaussian quantum states, pushing both fields into a new direction.

\end{abstract}

\maketitle

\section{Introduction}

Machine learning (ML) covers a wide range of algorithms and modelling tools with automated data-processing capabilities based on experience \cite{jordan2015machine,alpaydin2020introduction}. ML, with the prominent example of neural networks, has proven successful for tackling practical processing tasks that are unsuitable for conventional computer algorithms \cite{krizhevsky2012imagenet,hinton2012deep,lecun2015deep,mnih2015human,vinyals2019grandmaster}. With the deployment of ML algorithms, their limitations and inefficiencies when running on top of conventional computing hardware arise both in terms of power consumption and computing speed \cite{xu2018scaling}. The demand for an increased efficiency is currently fueling the field of unconventional computing, which aims at developing  hardware and algorithms that go beyond the traditional von Neumann architectures \cite{linn2012beyond,merolla2014million,zidan2018future}. Recent  extensions of neural networks and other ML techniques based on quantum systems  \cite{schuld2014quest,wittek2014quantum,dunjko2018machine,steinbrecher2019quantum,biamonte2017quantum} aim to offer and identify novel capabilities\cite{torrontegui2019unitary,beer2020training,killoran2019continuous,amin2018quantum,adhikary2020supervised,liu2020echo,shao2020quantum}. In this context, reservoir computing (RC) is a machine learning paradigm that is amenable to unconventional hardware-based approaches in the classical domain, e.g. in photonics \cite{brunner2013parallel,duport2012all,vandoorne2014experimental,larger2017high,van2017advances,brunner2019photonic,dong2019optical} and spintronics \cite{torrejon2017neuromorphic,nakane2018reservoir},  and has the potential to be extended to the quantum regime \cite{fujii2017harnessing,negoro2018machine,nakajima2019boosting,chen2019learning,ghosh2019quantum,liu2020echo,chen2020temporal,kutvonen2020optimizing,marcucci2020programming,govia2020quantum,ghosh2020universal,ghosh2020reconstructing,martinez2020information}.

RC exploits the dynamics of a non-linear system---the reservoir---for information processing of time dependent inputs \cite{verstraeten2007experimental,lukovsevivcius2009reservoir}. RC has its roots in the discovery that in recurrent neural networks, i.e. neural networks with an internal state, it is sufficient to only train the connections leading to the final output layer, avoiding optimization difficulties well-known in neural networks, without any apparent loss in computational power \cite{maass2002real,jaeger2004harnessing}. In practice, reservoir computers have achieved state-of-the-art performance in tasks such as continuous speech recognition \cite{triefenbach2014large} and nonlinear time series prediction \cite{pathak2018model} thanks to their intrinsic memory \cite{jaeger2001echo}.

Here, we put forward the framework for RC with continuous-variable quantum systems, in bosonic reservoirs given by harmonic networks with Gaussian states. This proposal could be implemented in tailored multimode optical parametric processes \cite{nokkala2018reconfigurable}, that already realize versatile and large entanglement networks in several experimental platforms \cite{Cai17,Chen14,Asavanant19}. Multimode quantum states of few modes have also been recently implemented in superconducting and optomechanical platforms \cite{Chang19sup,Lenzini18int,Nielsen17optmec,kollar2019hyperbolic}. An optical implementation would have intrinsic resilience to decoherence even at room temperature and allow to easily read-out a portion of the optical signal for output processing, including (direct or indirect) measurements. In any platform, an advantage of RC is that the reservoir
does not rely on fine tuning; we will actually consider random networks.
Therefore in the quantum regime these systems are well suited for NISQ (noisy intermediate-scale quantum) technologies  \cite{preskill2018quantum}.
The general theoretical framework we introduce here explores the utility of the method spanning across classical and quantum states and is  applicable to several physical reservoirs and any temporal task.

The restriction to Gaussian dynamics of linear oscillators brings the model within reach of state-of-the-art experimental platforms, but one might expect it to have very modest information processing capabilities. Indeed, previous proposals for 
neural networks realized with continuous variable quantum systems have either used non-Gaussian gates \cite{killoran2019continuous} or nonlinear oscillators \cite{govia2020quantum} as a source of nonlinearity, a crucial resource for nontrivial information processing. Surprisingly, we find that in order to achieve universal RC,  Gaussian resources and linear oscillators are enough. Non-Gaussian gates are instead necessary for universal quantum computing  in CV \cite{lloyd1999quantum}, constituting an unsolved engineering challenge so far \cite{GKP01,Yukawa13,Arzani17,Sabapathy18}. The concept of universality in different fields and applications  depends on the scope. In the context of quantum computing it refers to the ability to reproduce any unitary transformation \cite{lloyd1999quantum,nielsen-chuang}, in feed-forward neural networks it characterizes the possibility to approximate any 
continuous function \cite{cybenko1989approximation,hornik1989multilayer}, and in RC it identifies the ability to approximate as accurately as one wants so called fading memory functions \cite{maass2004computational,grigoryeva2018echo}.
We show that a quantum or classical linear network with only Gaussian states provides an analogue system (not based on gates or circuits) that is universal for RC when different encodings are exploited. Universality can be achieved in Gaussian RC by combining the outputs from a finite number of different networks (which can be thought of as a large network with many components) with a polynomial readout function. 

Besides constituting a universal class for RC, this linear quantum network is also versatile for solving different temporal tasks. Even in the case of a readout function which is linear in the observables of a fixed single component network, the overall input-output map can be tuned from fully linear to highly nonlinear by exploiting the input encoding of Gaussian quantum states. Thanks to this tunability, a fixed network can then adapt to the requirements of the task at hand, providing a further advantage to this proposal of RC. The generality of the performance analysis is achieved considering the information processing capacity (IPC) \cite{dambre2012information}, which allows for a task-independent assessment of time series processing and used here for the first time for a continuous-variable quantum system.

An important novelty of our RC scheme is the encoding of information in field fluctuations, or covariances, instead of in field intensities, or first moments, resulting in a significantly increase of the system’s IPC. However, we find that classical (thermal) fluctuations of a Gaussian system cannot provide any nonlinear memory. To solve the problem we take a step towards exploiting the quantumness of the system and propose a RC scheme that fully works with quantum fluctuations provided by encoding to squeezed vacuum. Indeed, this achieves simultaneously increased capacity, nonlinear memory and universal RC.

\section{Results}

In the following, after introducing RC with Gaussian states {(Sect.~\ref{sec:model})}, we demonstrate that this novel RC approach is universal {(Sect.~\ref{sec:universality})}. We then show numerical evidence of its performance and versatility enabled by encoding the input to different quantum states, as one can control the degree of nonlinearity of the reservoir by tuning the encoding (Sections~\ref{sec:nonlinearity} and \ref{sec:IPCQRC}). Finally, we illustrate the computational advantage when encoding the input to squeezed vacuum (Sect.~\ref{sec:IPC}).

\subsection{The model}\label{sec:model}

\begin{figure}[t]
                \includegraphics[trim=1cm 1cm 2cm 0cm,clip=true,width=0.45\textwidth]{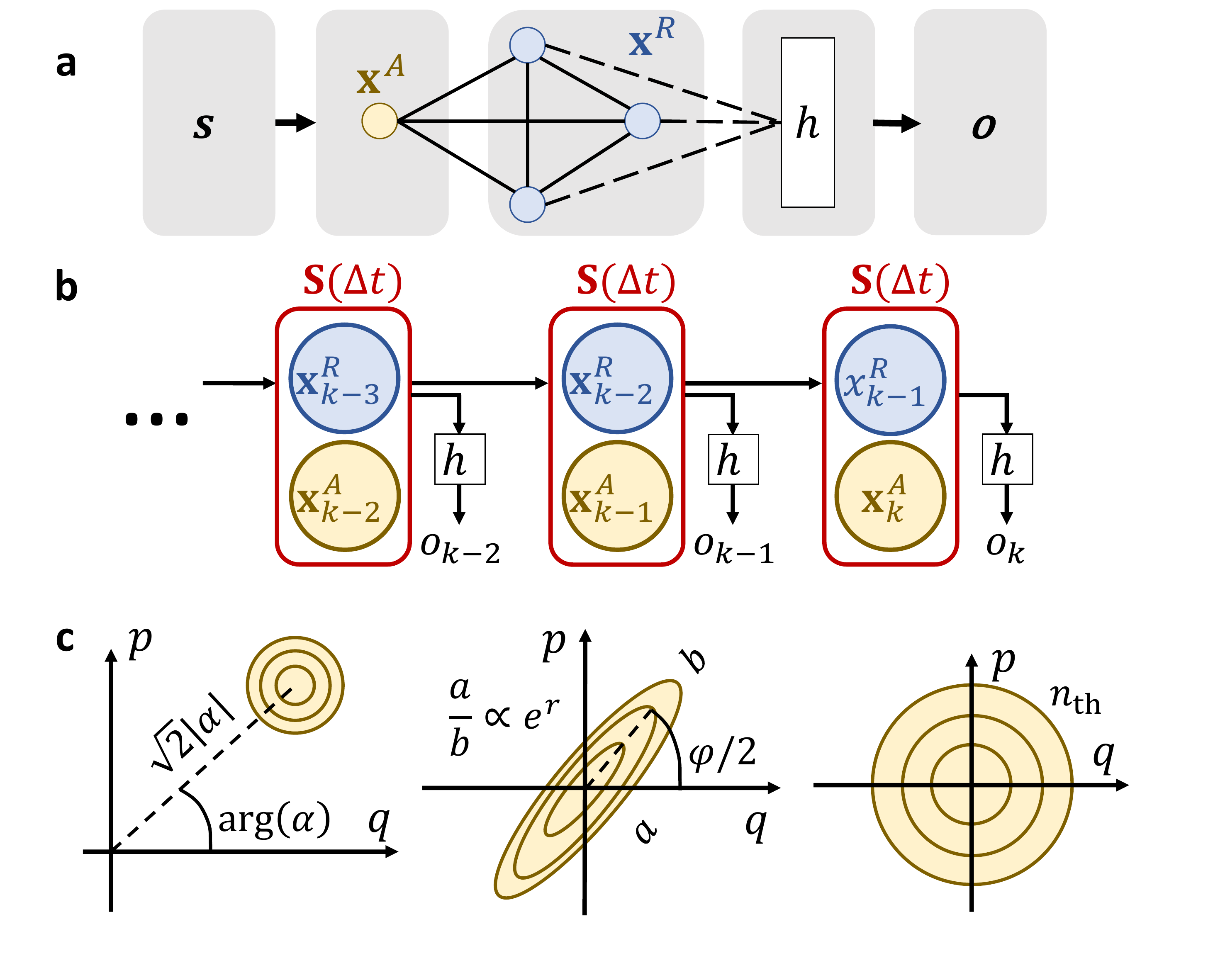}
            \caption{\label{fig:schematic}
\textbf{Reservoir computing scheme.} \textbf{a} The overall input-output map. The input sequence $\mathbf{s}$ is mapped to a sequence of ancillary single-mode Gaussian states. These states are injected one by one into a suitable fixed quantum harmonic oscillator network by sequentially resetting the state of the oscillator chosen as the ancilla, $\mathbf{x}^A$. The rest of the network---taken to be the reservoir---has operators $\mathbf{x}^R$. Network dynamics maps the ancillary states into reservoir states, which are mapped to elements of the output sequence $\mathbf{o}$ by a trained function $h$ of reservoir observables. \textbf{b} The corresponding circuit. The reservoir interacts with each ancillary state through a symplectic matrix $\mathbf{S}(\Delta t)$ induced by the network Hamiltonian $H$ during constant interaction time $\Delta t$. Output ($o_k$) at timestep $k$  is extracted before each new input. $\mathbf{x}^A_k$ are the ancillary operators conditioned on input $\mathbf{s}_k$ and $\mathbf{x}^R_k$ are the reservoir operators after processing this input. \textbf{c} Wigner quasiprobability distribution of ancilla encoding states in phase space. Input may be encoded in coherent states using amplitude $|\alpha|$ and phase $\mathrm{arg}(\alpha)$, or in squeezed states using squeezing parameter $r$ and phase of squeezing $\varphi$, or in thermal states using thermal excitations $n_{th}$.
}
      \end{figure}

We consider a linear network of $N$ interacting quantum harmonic oscillators as detailed in Methods \ref{sec:Hamiltonian}. The scheme for using this network for RC is shown in Fig.~\ref{fig:schematic}. The input sequence $\textbf{s}=\{\ldots,\textbf{s}_{k-1},\textbf{s}_k,\textbf{s}_{k+1},\ldots\}$, where $\textbf{s}_{k}\in \mathbb{R}^n$ represents each input vector and $k\in\mathbb{Z}$, is injected into the network by resetting at each timestep $k$ the state of one of the oscillators, called ancilla ($A$), accordingly.  As often done in RC, each input vector can be mapped to a single scalar in the ancilla through a function of the scalar product $\mathbf{v}^\top\cdot\mathbf{s}_k$ where $\mathbf{v}\in \mathbb{R}^n$. The rest of the network acts as the reservoir ($R$), and output is taken to be a function {$h$} of the reservoir observables before each input.

To express the dynamics, let $\mathbf{x}^\top=\{q_1,p_1,q_2,p_2,\ldots\}$ be the vector of network momentum and position operators and let $\mathbf{x}_k$ be the form of this vector at timestep $k$, after input $\mathbf{s}_k$ has been processed. We may take the ancilla to be the $N$th oscillator without a loss of generality. Let the time between inputs be $\Delta t$. Now operator vector $\mathbf{x}_k$ is related to $\mathbf{x}_{k-1}$ by
\begin{equation}
\mathbf{x}_k=\mathbf{S}(\Delta t)
\left(P_R\mathbf{x}_{k-1}\oplus\mathbf{x}_{k}^A\right),
\label{eq:networkdynamics}
\end{equation}
where $P_R$ drops the ancillary operators from $\mathbf{x}_{k-1}$ (reservoir projector, orthogonal to the ancilla vector) and $\mathbf{x}_{k}^A$ is the vector of ancillary operators conditioned on input $\mathbf{s}_k$, while $\mathbf{S}(\Delta t)\in\textrm{Sp}(2N,\mathbb{R})$ is the symplectic matrix induced by the Hamiltonian in Eq.~\eqref{eq:ham} (see Methods) and time $\Delta t$
(for more details on this formalism see, e.g., \cite{ferraro2005gaussian,adesso2014continuous,nokkala2017non}). The dynamics of reservoir operators
 $\mathbf{x}_k^R=P_R\mathbf{x}_{k}$ 
 is conveniently described dividing $\mathbf{S}(\Delta t)$ into blocks as
\begin{equation}
\mathbf{S}(\Delta t)=\begin{pmatrix}
\mathbf{A} & \mathbf{B} \\
\mathbf{C} & \mathbf{D} 
\end{pmatrix},
\label{eq:Sblocks}
\end{equation}
where $\mathbf{A}$ is $2(N-1)\times 2(N-1)$ and $\mathbf{D}$ is $2\times 2$. Now the formal definition of the proposed reservoir computer reads 
\begin{equation}
\begin{cases}
\mathbf{x}_k^R= \mathbf{A}\mathbf{x}_{k-1}^R+\mathbf{B}\mathbf{x}_{k}^A,\\
o_k=h(\mathbf{x}_k^R),
\end{cases}
\label{eq:qrc}
\end{equation}
where $h$ maps the reservoir operators to elements of the real output sequence  $\mathbf{o}=\{\ldots,o_{k-1},o_k,o_{k+1},\ldots\}$.

For Gaussian states, the full dynamics of the system conditioned by the sequential input injection is entailed in the first moments vector $\langle\mathbf{x}_k^R\rangle$ and covariance matrix $\sigma(\mathbf{x}_k^R)$. The values at step $0$, given a sequence of previous $m$ inputs  $\mathbf{s}=\{\mathbf{s}_{-m+1},\ldots,\mathbf{s}_{-1},\mathbf{s}_0\}$ encoded in the corresponding ancilla vectors, is obtained through repeated application of Eqs.~\eqref{eq:qrc} and reads
\begin{equation}
\begin{cases}
\begin{aligned}
\sigma(\mathbf{x}_0^R)&=\mathbf{A}^m\sigma(\mathbf{x}_{-m}^R)(\mathbf{A}^\top)^m\\&+\sum_{j=0}^{m-1}\mathbf{A}^{j}\mathbf{B}\sigma(\mathbf{x}_{-j}^A)\mathbf{B}^\top(\mathbf{A}^\top)^{j},
\end{aligned}
\\
\langle\mathbf{x}_0^R\rangle=\mathbf{A}^m\langle\mathbf{x}_{-m}^R\rangle+\sum_{j=0}^{m-1}\mathbf{A}^{j}\mathbf{B}\langle\mathbf{x}_{-j}^A\rangle,
\end{cases}
\label{eq:qrcchannel}
\end{equation}
where $\sigma(\mathbf{x}_{-m}^R)$ and $\langle\mathbf{x}_{-m}^R\rangle$ are the initial conditions, i.e. the initial state of the reservoir. This is the Gaussian channel for the reservoir conditioned on the input encoded in $\mathbf{x}^A$. Different Gaussian states of the ancilla can be addressed, such as coherent states, squeezed vacuum or thermal states  (see Fig.~\ref{fig:schematic}), respectively characterized by the complex displacement $\alpha$,
squeezing degree $r$ and phase $\varphi$,
and thermal excitations $n_{th}$ (see Methods \ref{ap:cov_matrix}, Eqs.~\eqref{eq:methods1}). Finally, the output is taken to be either linear or polynomial in either the elements of $\sigma(\mathbf{x}_k^R)$ or $\langle\mathbf{x}_k^R\rangle$. 
Observables can be be extracted with Gaussian measurements such as homodyne or heterodyne detection. We will next show that the introduced model not only satisfies the requirements for RC, but is notably even universal for RC even when restricting to specific input encodings.

\subsection{Universality for reservoir computing}\label{sec:universality}

To begin with, we show that instances of the model defined by  Eqs.~\eqref{eq:qrc} and the dependency of $\mathbf{x}_k^A$ on $\mathbf{s}_k$ can be used for RC, i.e. the dynamics conditioned by the input can be used for online time series processing by adjusting the coefficients of the polynomial defined by $h$ to get the desired output.

As explained in Methods \ref{sec:RCtheory}, the goal is more formally to reproduce a time-dependent function $f(t)=F[\{\ldots,\mathbf{s}_{t-2},\mathbf{s}_{t-1},\mathbf{s}_{t}\}]$, associated with given input $\mathbf{s}$ and functional $F$ from the space of inputs to reals. Consequently, we say that the system can be used for RC if there is a functional 
from the space of inputs to reals that is both \textit{a solution of Eqs.~\eqref{eq:qrc}} and \textit{sufficiently well-behaved} to facilitate learning of different tasks. These two requirements are addressed by the echo state property (ESP) \cite{jaeger2001echo} and the fading memory property (FMP) \cite{boyd1985fading}, respectively. The associated functions are called fading memory functions. In essence, a reservoir has ESP if and only if it realizes a fixed map from the input space to reservoir state space---unchanged by the reservoir initial conditions---while FMP means that to get similar outputs it is enough to use inputs similar in recent past---which provides, e.g., robustness to small past changes in input. The two are closely related and in particular both of them imply that the reservoir state will eventually become completely determined by the input history; in other words forgetting the initial state is a necessary condition for ESP and FMP.

Looking at Eqs.~\eqref{eq:qrcchannel}, it is readily seen that the model will become independent of the initial conditions at the limit $m\to\infty$ of a left infinite input sequence if and only if $\rho(\mathbf{A})<1$, where $\rho(\mathbf{A})$ is the spectral radius of matrix $\mathbf{A}$. Therefore, $\rho(\mathbf{A})<1$ is a necessary condition for having ESP and FMP. The following lemma (proven in Supplementary Information) states that it is also sufficient when we introduce the mild constraint of working with uniformly bounded subsets of the full input space, i.e. there is a constant that upper bounds $\left\lVert\mathbf{s}_k\right\rVert$ for all $k$ in the past.
\begin{lemma}\label{lemma1}
Suppose the input sequence $\mathbf{s}$ is uniformly bounded. Let ancilla parameters be continuous in input and let $h$ be a polynomial of the elements of $\sigma(\mathbf{x}_k^R)$ or $\langle\mathbf{x}_k^R\rangle$. The corresponding reservoir system has both ESP and FMP if and only if the matrix $\mathbf{A}$ in Eqs.~\eqref{eq:qrc} fulfills $\rho(\mathbf{A})<1$.
\end{lemma}
\noindent This is the sought condition for RC with harmonic networks, either classical or quantum. Importantly, it allows to discriminate useful reservoirs by simple inspection of the parameters of the network through the spectral radius of $\mathbf{A}$.

We now turn our attention to universality. The final requirement to fulfill is separability, which means that for any pair of different time series there is an instance of the model that can tell them apart. Then the class of systems defined by Eqs.~\eqref{eq:qrc} is universal \cite{grigoryeva2018universal,grigoryeva2018echo} in the following sense. For any element $F$ in a class of fading memory functionals that will be given shortly, there exists a finite set of functionals realized by our model that can be combined to approximate $F$ up to any desired accuracy. Physically, such combinations can be realized by combining the outputs of many instances of the model with a polynomial function. Mathematically, this amounts to constructing the polynomial algebra of functionals.

The next theorem (of which we give a simplified version here and a full version in the Supplementary Information) summarizes our analysis of the model described in Eqs.~\eqref{eq:qrc}.

\begin{theorem*}[\textbf{simplified}]
Given instances of reservoir systems governed by Eqs.~\eqref{eq:qrc} with a given $\Delta t$ and for $\rho(A)<1$, hence providing a family $\mathcal{Q}$ of fading memory functionals, the polynomial algebra of these functionals has separability. This holds also for the subfamilies $\mathcal{Q}_{\text{thermal}}$, $\mathcal{Q}_{\text{squeezed}}$ and $\mathcal{Q}_{\text{phase}}$, that correspond to thermal, squeezed and phase encoding respectively. As any causal, time-invariant fading memory functional $F$ can be uniformly approximated by its elements, the reservoir family of Eqs.~\eqref{eq:qrc} with the specified constraint is universal.
\end{theorem*}

We sketch the main ingredients of the proof. Since the model admits arbitrarily small values of $\rho(\mathbf{A})$, there are instances where $\rho(\mathbf{A})<1$; by virtue of Lemma 1, $\mathcal{Q}$ can therefore be constructed and it is not empty. We show that the associated algebra has separability. Since the space of inputs is taken to be uniformly bounded, we may invoke the Stone-Weierstrass Theorem \cite{dieudonne2011foundations} and the claim follows. Full proof and additional details are in Supplementary Information.

We note that unlike ESP and FMP, separability depends explicitly on the input encoding. In Supplementary Information we show separability for three different encodings of the input to elements of $\sigma(\mathbf{x}_k^A)$: thermal ($n_{th}$), squeezing strength ($r$) and phase of squeezing ($\varphi$). It should be pointed out that separability (and hence, universality) could be shown also for first moments encoding in a similar manner.

\subsection{Controlling performance with input encoding}\label{sec:nonlinearity}

The here derived Universality Theorem guarantees that for any temporal task, there is a finite set of reservoirs and readouts that can perform it arbitrarily well when combined. Let us now assume a somewhat more practical point of view: we possess a given generic network, and we attempt to succeed in different tasks by training the output function $h$ to minimize the squared error between output $\mathbf{o}$ and target output. For simplicity, we will also take inputs to be sequences of real numbers, rather than sequences of vectors---we stress that universality holds for both. Beyond universality of the whole class of Gaussian reservoirs, what is the performance and versatility of a generic one?

First of all, we will address how to single out instances with good memory. As pointed out earlier, memory is provided by the dependency of the reservoir observables on the input sequence. Informally speaking, reservoirs with good memory can reproduce a wide range of functions of the input and therefore learn many different tasks. Furthermore, to be useful a reservoir should possess nonlinear memory, since this allows the offloading of nontrivial transformations of the input to the reservoir. Then nonlinear time series processing can be carried out while keeping the readout linear, which simplifies training and reduces the overhead from evaluating the trained function.

Memory is strongly connected to FMP; in fact, a recent general result concerning reservoirs processing discrete-time data is that under certain mild conditions, FMP guarantees that the total memory of a reservoir---bounded by the number of linearly independent observables used to form the output---is as large as possible \cite{dambre2012information}. Consequently, all instances that satisfy the spectral radius condition of Lemma $1$ have maximal memory in this sense. Indeed with Lemma $1$ the condition for FMP is straightforward to check. Furthermore, we find that reservoir observables seem to be independent as long as $\mathbf{L}$ does not have special symmetries---as a matter of fact, numerical evidence suggests a highly symmetric network such as a completely connected network with uniform frequencies and weights never satisfies $\rho(\mathbf{A})<1$. Having FMP says nothing about what kind of functions the memory consists of, however.

\begin{figure}[t]
                \includegraphics[trim=0cm 0.1cm 0.2cm 0cm,clip=true,width=0.45\textwidth]{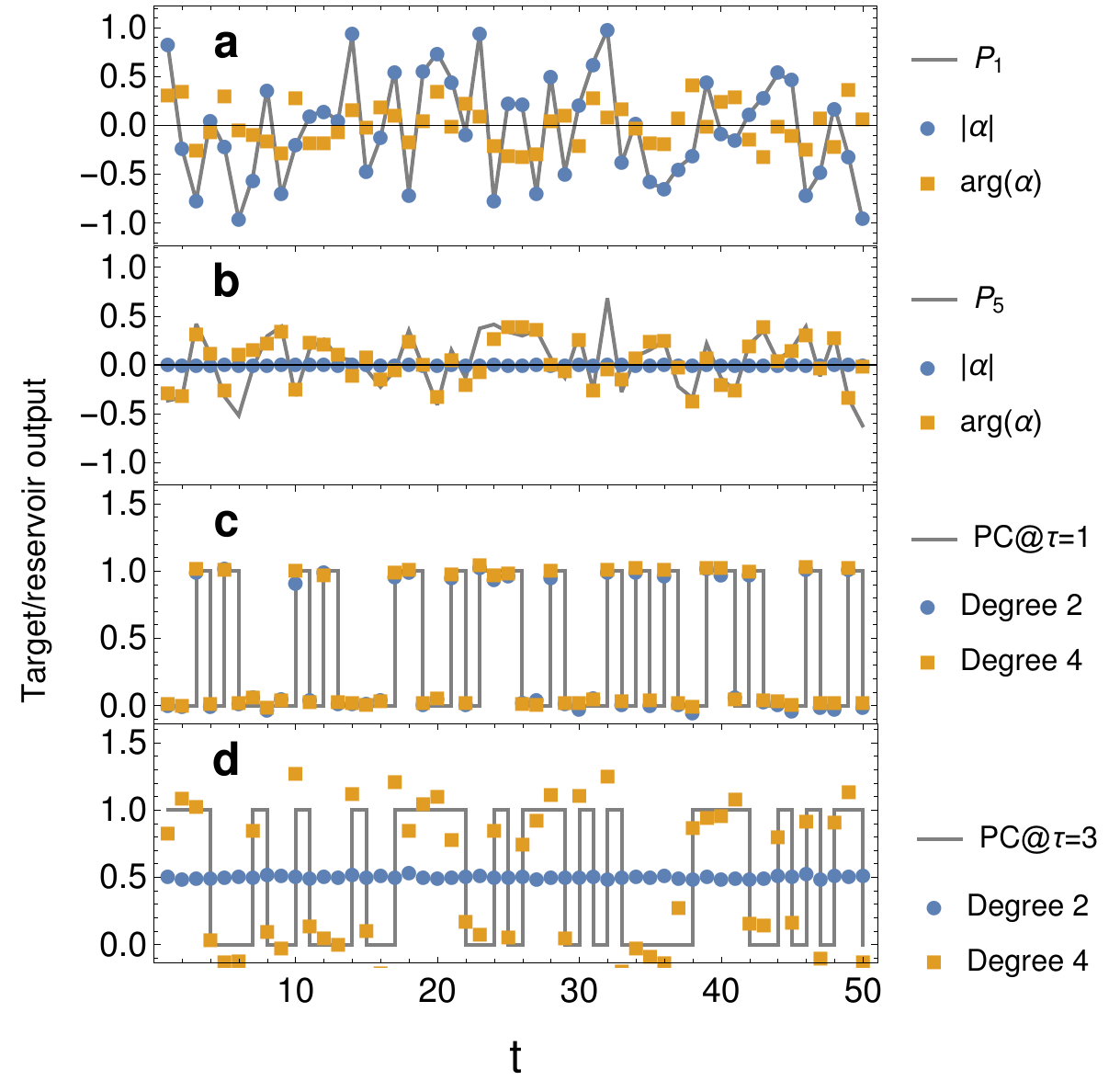}
            \caption{\label{fig:predictcomparison}
\textbf{Nonlinear information processing with a generic reservoir.} In panels \textbf{a} and \textbf{b} the targets are $(P_1)_k=s_k$ and $(P_5)_k=(15s_k-70s_k^3+63s_k^5)/8$ while $h$ is a trained linear function of $\langle\mathbf{x}_k^R\rangle$. Encoding the input to either magnitude of displacement $|\alpha|$ or phase $\mathrm{arg}(\alpha)$ of the coherent state is compared. The former is able to reproduce only the linear target $P_1$ while the latter
has good performance with $P_5$, confirming that some nonlinear tasks are possible with linear $h$. In panels \textbf{c} and \textbf{d} we fix the encoding to $|\alpha|$ and consider the parity check task (PC@$\tau=1,3$) which requires products of the inputs at different delays $\tau$. These terms can be introduced by a polynomial $h$ (degree$=2,4$); increasing its degree allows the task to be reproduced at increasingly long delays. In all cases a network of $N=8$ oscillators is used and the reservoir output is compared to target for $50$ time steps after training (see Methods \ref{sec:usednetworks} and \ref{sec:training} for details).}
      \end{figure}

Is there nonlinear memory? It is evident that the first of Eqs.~\eqref{eq:qrc} is linear in reservoir and ancilla operators, but the encoding is not necessarily linear because of the way ancilla operators $\mathbf{x}_{k}^A$ depend on the input. For single-mode Gaussian states (see Eqs.~\eqref{eq:methods1} in Methods \ref{ap:cov_matrix}), it can be seen that the reservoir state is nonlinear in input when encoding to either magnitude $r$ or phase $\varphi$ of squeezing, or the phase of displacement $\mathrm{arg}(\alpha)$. Otherwise, that is for encoding in coherent states amplitude or thermal states average energy, it is linear (see Eqs.~\eqref{eq:methods2} in Methods \ref{ap:cov_matrix}). This implies that the input encoding is the only source of nonlinear memory when the readout is linear---nonlinearity comes from pre-processing the input to ancilla states, not from an activation function which is the conventional source of nonlinearity in RC.

The performance of Gaussian RC can be assessed considering different scenarios. For the remainder of this work we fix the network size to $N=8$ oscillators and form the output using the reservoir observables and a bias term; see Methods \ref{sec:usednetworks} and \ref{sec:training} for details. We consider nonlinear tasks in Fig.~\ref{fig:predictcomparison}. In panels \textbf{a} and \textbf{b} we take  the output function  $h$ to be a linear function of $\langle\mathbf{x}_{k}^R\rangle$ and inputs $s_k$ to be uniformly distributed in $[-1,1]$, and consider two different encodings of the input into the ancilla $\langle\mathbf{x}_{k}^A\rangle$, as the amplitude and phase of coherent states. Setting $|\alpha|\to s_k+1$ and phase to a fixed value $\mathrm{arg}(\alpha)\to 0$ leads to fully linear memory, which leads to good performance in the linear task of panel \textbf{a} only. In contrast, setting $|\alpha|\to 1$ and encoding the input to phase as $\mathrm{arg}(\alpha)\to 2\pi s_k$ leads to good performance in the nonlinear task shown in panel \textbf{b} and limited success in the linear one \textbf{a}.

Nonlinearity of reservoir memory is not without limitations since $\langle\mathbf{x}_{k}^R\rangle$ does not depend on products of the form $s_k s_j \cdots s_l$ where at least some of the indices are unequal, i.e. on products of inputs at different delays, as can be seen from Eq.~\eqref{eq:methods2}. This is a direct consequence of the linearity of reservoir dynamics. When $h$ is linear in $\langle\mathbf{x}_{k}^R\rangle$ the output will also be independent of these product terms, hindering performance in any temporal task requiring them. While Universality Theorem implies the existence of a set of reservoirs for any task, we will show that even a single generic reservoir can be sufficient when nonlinearity is introduced at the level of readout, at the cost of relatively more involved training.

To illustrate the nontrivial information processing power of a single generic reservoir, we consider the parity check task \cite{bertschinger2004real} (Fig.~\ref{fig:predictcomparison} \textbf{c} and \textbf{d}), defined as
\begin{equation}
(\mathrm{PC}(\tau))_k=\mathrm{mod}\left(\sum_{l=0}^{\tau}s_{k-l},2 \right)
\label{eq:PCtask}
\end{equation}
where $s_k\in\{0,1\}$; the target output is $0$ if the sum of $\tau+1$ most recent inputs is even and $1$ otherwise. It can be shown that $(\mathrm{PC}(\tau))_k$ coincides with a sum of products of inputs at different delays for binary input considered here. When encoding in a coherent state amplitude ($|\alpha|\to s_k$, $\mathrm{arg}(\alpha)\to 0$) for readout function $h$ polynomial in $\langle\mathbf{x}_{k}^R\rangle$, the results show that increasing the polynomial degree $d$ of the function $h$ allows to solve the task for higher delays $\tau$. In particular, we find that the reservoir can solve the well-known XOR problem for nonlinearity degrees $d\geq 2$, which coincides with the parity check at $\tau=1$. The parity check at $\tau =3$ works, in turn, for $d\geq 4$.

\subsection{Information processing capacity}\label{sec:IPCQRC}

Besides providing nonlinearity, input encoding also facilitates a versatile tuning of the linear and nonlinear memory contributions. To demonstrate this, we consider how input encoding affects the degree of nonlinear functions that the reservoir can approximate, as quantified by the information processing capacity (IPC) \cite{dambre2012information} of the reservoir. The IPC generalizes the linear memory capacity \cite{jaeger2002short} often considered in RC to both linear and nonlinear functions of the input. Even if its numerical evaluation is rather demanding, it has the clear advantage to provide a broad assessment of the features of RC, beyond the specificity of different tasks. 

We may define the IPC as follows. Let $X$ be a fixed reservoir, $z$ a function of a finite number of past inputs and let $h$ be $linear$ in the observables of $X$. Suppose the reservoir is run with two sequences $\mathbf{s}'$ and $\mathbf{s}$ of random inputs drawn independently from some fixed probability distribution $p(s)$. The first sequence $\mathbf{s}'$ is used to initialize the reservoir; observables are recorded only when the rest of the inputs $\mathbf{s}$ are encoded. The capacity of the reservoir $X$ to reconstruct $z$ given $\mathbf{s}'$ and $\mathbf{s}$ is defined to be 
\begin{equation}
C_{\mathbf{s}',\mathbf{s}}(X,z)=1-\dfrac{\mathrm{min}_h\sum_k(z_k-o_k)^2}{\sum_k z_k^2}
\label{eq:capacity}
\end{equation}
where the sums are over timesteps $k$ after initialization, each $z_k$ is induced by the function $z$ to be reconstructed and the input, and we consider the $h$ that minimizes the squared error in the numerator. The maximal memory mentioned earlier may be formalized in terms of capacities: under the conditions of Theorem $7$ in Ref.~\onlinecite{dambre2012information}, the sum of capacities $C_{\mathbf{s}',\mathbf{s}}(X,z)$ over different functions $z$ is upper bounded by the number of linearly independent observables used by $h$, with the bound saturated if $X$ has FMP. This also implies that finite systems have finite memory, and in particular increasing the nonlinearity of the system inevitably decreases the linear memory \cite{ganguli2008memory,dambre2012information}. Importantly, infinite sequences $\mathbf{s}'$, $\mathbf{s}$ and a set of functions that form a complete orthogonal system w.r.t. $p(s)$ are required by the theorem; shown results are numerical estimates. We consistently take $p(s)$ to be the uniform distribution in $[-1,1]$; examples of functions $z$ orthogonal w.r.t. this $p(s)$ include Legendre polynomials $P_1$ and $P_5$ appearing in Fig.~\ref{fig:predictcomparison}, as well as their delayed counterparts. Further details are given in Methods \ref{sec:IPCmethods}.
 
\begin{figure}[t]
                \includegraphics[trim=0cm 0cm 0cm 0cm,clip=true,width=0.45\textwidth]{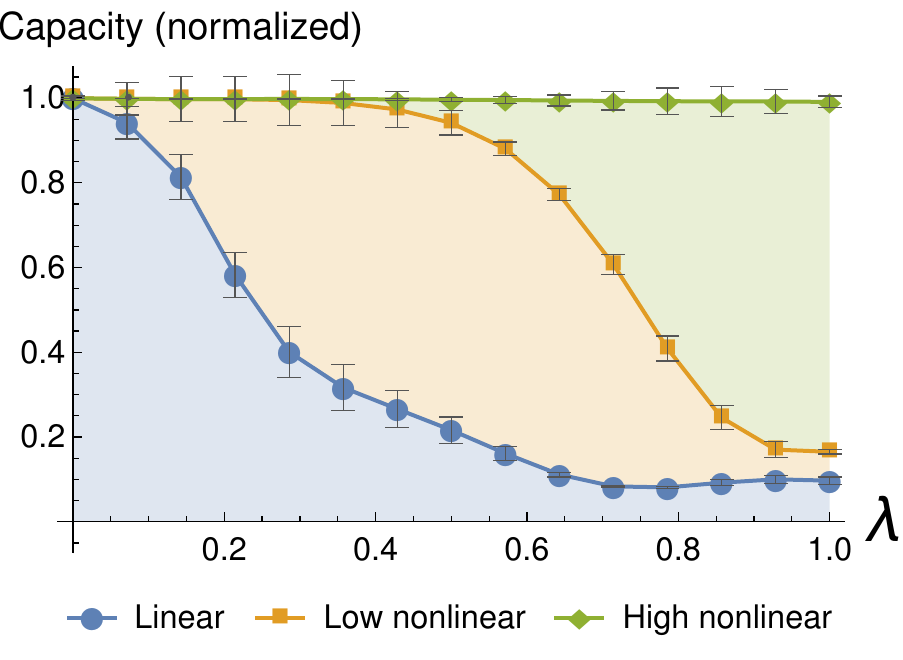}
            \caption{\label{fig:NLtuning}
\textbf{Control of nonlinearity of reservoir memory via input encoding.} Here we set $|\alpha|\to (1-\lambda)(s_k+1)+\lambda$, $\mathrm{arg}(\alpha)\to 2\pi\lambda s_k$ where the input is $s_k\in[-1,1]$. Reservoir memory is measured using information processing capacity, which quantifies the ability of the reservoir to reconstruct functions of the input at different delays.
The figure shows how the relative contributions from linear and nonlinear functions to the normalized total capacity can be controlled with $\lambda$. Nonlinear contributions are further divided to degrees $2$ and $3$ (low nonlinear) and higher (high nonlinear). For $\lambda=0$ the encoding is strictly to $|\alpha|$, leading to linear information processing, while at $\lambda=1$ only $\mathrm{arg}(\alpha)$ depends on the input, leading to most of the capacity to come from functions of the input with degree at least $4$. All results are averages over $100$ random reservoirs and error bars show the standard deviation.
}
      \end{figure}

We consider the breakdown of the normalized total capacity to linear (covering functions $z$ with degree $0$ or $1$), nonlinear (degrees $2$-$3$) and highly nonlinear (degree $4$ or higher) regimes in Fig.~\ref{fig:NLtuning}. We take $h$ to be a linear function of $\langle\mathbf{x}_{k}^R\rangle$ and  address the capability to have Gaussian RC operating with different linear and non-linear capabilities by varying the input encoding
into a coherent ancillary state from amplitude to phase $|\alpha|\to (1-\lambda)(s_k+1)+\lambda$, $\mathrm{arg}(\alpha)\to 2\pi\lambda s_k$ where $s_k\in[-1,1]$; this is a convex combination of the two encodings used in panels \textbf{a} and \textbf{b} of Fig.~\ref{fig:predictcomparison}. As can be seen in Fig.~\ref{fig:NLtuning}, adjusting $\lambda$ allows one to move from fully linear (for amplitude encoding) to highly nonlinear (for phase) information processing, which can be exploited for a versatile tuning of the reservoir to the task at hand. Remarkably, this can be done without changing neither the parameters of the Hamiltonian \eqref{eq:ham} (that is, the reservoir system) nor the observables extracted as output in $h$. The earlier mentioned trade-off between linear and nonlinear memory keeps the total memory bounded, however Lemma 1 ensures that this bound is saturated for all values of $\lambda$.

\subsection{From intensities to field fluctuations and RC with quantum resources}\label{sec:IPC}

Previously we considered coherent states for the ancilla, encoding the input to $|\alpha|$ and $\mathrm{arg}(\alpha)$. In the limit of large amplitudes $|\alpha|\gg 1$, coherent states allow for a classical description of the harmonic network, with field operator expectation values corresponding, for instance, to classical laser fields \cite{schrodinger1926stetige,glauber1963coherent}. This realization of RC would be impossible in the limit of vanishing fields where $|\alpha|\to 0$ since then $\langle\mathbf{x}_{k}^R\rangle$ becomes independent of input. Here we put forward the idea of harnessing instead the fluctuations encoded in the covariance matrix $\sigma(\mathbf{x}_k^R)$, which also increases the amount of linearly independent reservoir observables. Even if only a subset of them can be directly monitored, the rest will play the role of hidden nodes that may be chosen to contribute as independent computational nodes, and have already been suggested to be a source of computational advantage in quantum RC in spin systems \cite{fujii2017harnessing}.  Here we analyze the full IPC of the system including all observables.

In the classical regime, thermal fluctuations could be used, corresponding to \textit{thermal states} (encoding to $n_{th}$) for the ancilla. While representing an original approach, it will be seen that these classical fluctuations will provide only linear memory and as such cannot be used to solve nontrivial temporal tasks without relying on external sources of nonlinearity. We propose to solve this problem by relying instead on quantum fluctuations provided by encoding the input to \textit{squeezed vacuum} (encoding to $r$, $\varphi$), i.e. by adopting a quantum approach to RC with Gaussian states. By virtue of the Universality Theorem, either thermal or squeezed states encoding can be used for universal RC.

\begin{figure}[t]
                \includegraphics[trim=0cm 0cm 0cm 0cm,clip=true,width=0.45\textwidth]{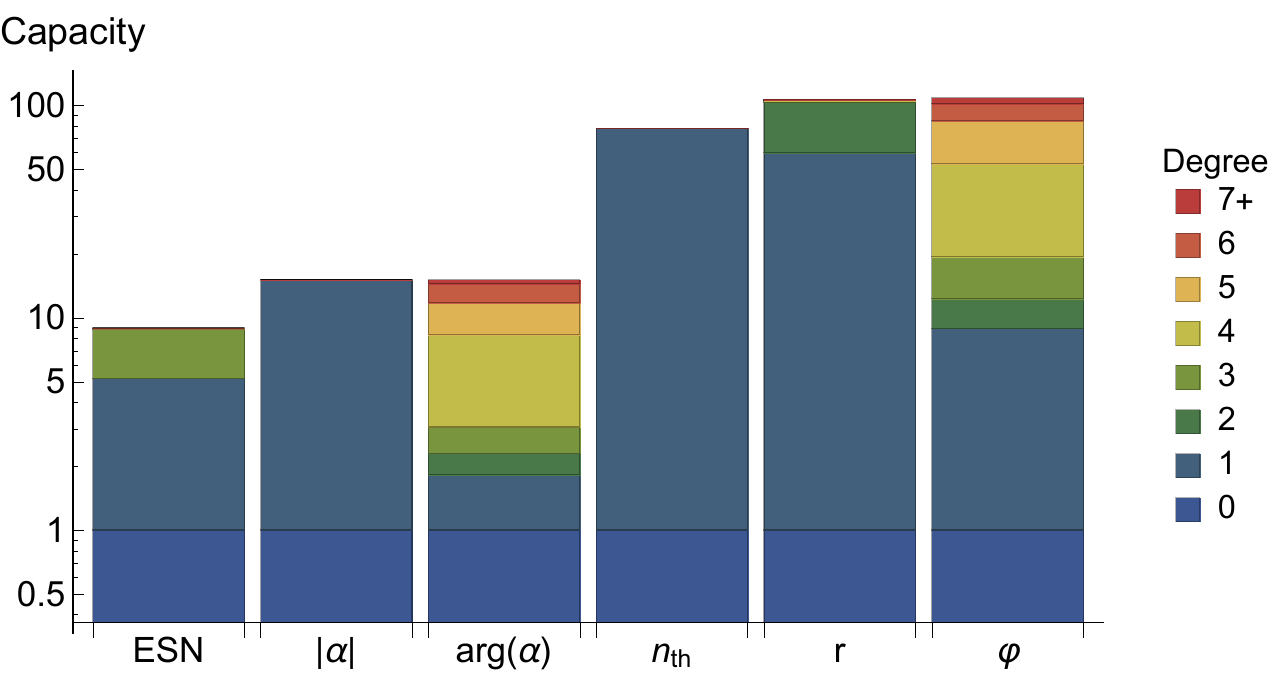}
            \caption{\label{fig:IPC}
\textbf{Histogram bars showing the information processing capacity for different input encodings.} Capacity of an echo state network (ESN) with as many neurons (8) as there are oscillators in the harmonic network is shown for comparison. The output is a function of first moments $\langle\mathbf{x}_k^R\rangle$ when encoding to either magnitude $|\alpha|$ or phase $\mathrm{arg}(\alpha)$ of displacement. Three different ways to encode the input in the limit $|\alpha|\to 0$ are shown; for them the output is a function of the elements of $\sigma(\mathbf{x}_k^R)$, with the input being encoded either to thermal excitations $n_{th}$, squeezing strength $r$ or angle $\varphi$.
}
\end{figure}

We compare the classical and quantum approaches in Fig.~\ref{fig:IPC}, where we show how the capacity of the system of 7 reservoir oscillators and the ancilla gets distributed for the aforementioned encodings. For comparison, we also include the case of an echo state network (ESN, see Methods \ref{sec:ESN}) consisting of 8 neurons, a classical reservoir computer based on a recurrent neural network \cite{jaeger2001echo}. We separate the contributions to total capacity according to degree to appreciate the linear memory capacity (degree 1) and nonlinear capacity and take $s_k$ to be uniformly distributed in $[-1,1]$. 
The total capacity, as displayed in Fig.~\ref{fig:IPC}, allows to visualize the clear advantage of the oscillators network over the ESN, as well as the particularities of the different input encodings.

The cases $|\alpha|$ and $\mathrm{arg}(\alpha)$ in Fig.~\ref{fig:IPC} correspond to Fig.~\ref{fig:NLtuning} for $\lambda=0$ and $\lambda=1$, respectively. For them, we take $h$ to be linear in $\langle\mathbf{x}_{k}^R\rangle$. The total capacity for ESN and harmonic networks with $|\alpha|$ and $\mathrm{arg}(\alpha)$ encoding differ by a factor 2 because the neurons of ESN each provide a single observable, while two quadratures are considered for reservoir oscillators. Comparing capacities, ESN provides more nonlinear memory than amplitude encoding of coherent states $|\alpha|$, but
for phase encoding   we can see significant nonlinear contributions to total capacity from higher degrees, up, e.g., to 7, in spite of the small size of the harmonic network.

Let us now reduce $|\alpha|\to 0$. In this case, we show in Fig.~\ref{fig:IPC} that classical thermal fluctuations provided by setting $n_{th}\to s_k+1$ and taking $h$ to be a linear function of $\sigma(\mathbf{x}_k^R)$ increase the capacity significantly, as the number of available observables is proportional to the square of the number of oscillators (see Methods \ref{sec:training}). For thermal encoding, the capacity does not reach the total capacity bound, limited by dependencies between some of the observables. 
Furthermore,  there is only linear memory. 
As shown in Fig.~\ref{fig:IPC}, considering instead quantum fluctuations gives both a quadratic capacity increase and nonlinear memory. Setting $r\to s_k+1$, $\varphi\to 0$ has a ratio of linear to nonlinear memory somewhat similar to the ESN, while setting $r\to 1$, $\varphi\to 2\pi s_k$ gives significant amounts of nonlinear memory (last bar in Fig.~\ref{fig:IPC}), displaying the advantage of encoding in squeezed vacuum.

\section{Discussion\label{sec:conclusions}}

In this work we have established that even with the limited resources offered by  linear quantum optics platforms, Gaussian measurements, and a restricted class of input Gaussian states, 
universal reservoir computing is achieved. Recently, universality of RC has been proved in the full class of classical linear systems  \cite{grigoryeva2018universal}, in echo-state networks \cite{grigoryeva2018echo}, and, on the quantum side, for the spins map of Ref. \onlinecite{chen2020temporal}. Here we go beyond these results demonstrating that universality holds in continuous variables with Gaussian operations and restricting to some states, like thermal or squeezed vacuum, providing the foundations for RC in linear optics platforms, classical or quantum, with minimal resources \cite{weedbrook2012gaussian,lami2018gaussian}.

We find that performance is not limited by the reduced resources in place, as well as a clear improvement when information is encoded in quantum fluctuations of the field, providing superior capacity.
This is displayed in the quadratic growth of IPC with system size, as opposed to linear in the case of first moments or standard ESNs. Versatility  is also achieved without modifying the network reservoir, as input encoding leads to easily tunable  nonlinear memory---already proven to be beneficial to tackle different computational tasks in classical RC \cite{soriano2015minimal,araujo2020role}. Non-classical resources in Gaussian systems originate from squeezing and this is shown to crucially  provide nonlinear memory necessary for achieving nontrivial RC with a linear readout clearly beyond the performance achieved with thermal fluctuations, which are classical.

RC realized with physical systems is a new and rapidly developing field \cite{tanaka2019recent}, with the extension to quantum regime proposed in a handful of pioneering works \cite{fujii2017harnessing,nakajima2019boosting,chen2019learning,negoro2018machine,chen2020temporal,govia2020quantum}.  
Our proposal can be adapted to many classical  and quantum platforms modeled by continuous variables in linear networks \cite{vandoorne2014experimental,nokkala2018reconfigurable,Cai17,Chang19sup,Masada15cont,Lenzini18int,Takeda2019,Nielsen17optmec,kollar2019hyperbolic}. 
As an example, any harmonic network described by Eq.~$(1)$ can be implemented in optical parametric processes pumped by optical frequency combs, based on interacting spectral or temporal modes. The network size and structure are controlled by shaping the pump and multimode measurements, allowing in principle to create any network within a fixed optical architecture \cite{nokkala2018reconfigurable}. Lemma 1  provides a simple and mild condition to identify networks with echo-state and fading memory properties in this reconfigurable platform. Even if the full numerical analysis of the IPC here is limited to $8$ nodes, larger networks have already been reported \cite{Cai17}. These linear quantum optics platforms are already well-established for measurement based quantum computing \cite{Cai17,Chen14,Asavanant19}, have intrinsic resilience to decoherence and high potential for scalability. Our study reveals their potential   as  promising physical systems for quantum RC.


The proposed scheme assumes Gaussian measurements as homodyne or heterodyne detection \cite{serafini}. A proof of principle realization of this protocol would require several copies of the experiment \cite{fujii2017harnessing,negoro2018machine,chen2020temporal} at each time. Otherwise, when monitoring the system output in order to exploit RC for temporal tasks,  the back-action effect of quantum measurement needs to be taken into account. 
Even if the system performance is robust up to some level of classical noise and the effective dissipation introduced by measurement turns out to be beneficial for the fading memory, the design of platforms including measurement protocols is a question open to further research. In optical implementations, a clear advantage is the possibility to exploit arbitrary-node manipulation via beam splitters operations and feedback loops, as in recent experiments \cite{Cai17,Takeda2019}. Measurements may also be a potential source of computational power introducing nonlinearity.  As examples, even for Gaussian states, non-Gaussian measurements allow to reproduce any unitary transformation \cite{menicucci2006universal} or to make a problem intractable for classical computers, as in Boson sampling \cite{Hamilton17}. 

The most direct advantage when considering RC with quantum systems is the possibility to access a larger Hilbert space \cite{fujii2017harnessing}. With the minimal resources of a Gaussian model, this  leads to a quadratic advantage. An interesting perspective is to explore a potential superior performance based on non-Gaussian resources \cite{Lvovsky20,Albarelli18,Takagi18}, to achieve exponential scaling of the total processing capacity in the quantum regime and a genuine quantum advantage in terms of computational complexity \cite{menicucci2006universal,Gu09}.   
A further possibility is to embed the (quantum) reservoir within the quantum system to be processed in order to explore quantum RC in quantum tasks, providing a quantum-quantum machine learning (in the data to be analyzed and in the RC platform, respectively).  Finally, the extension of the learning theory to quantum output is still missing; tasks of interests could include training the reservoir to simulate given quantum circuits, for instance. Static versions of such tasks carried out in feed-forward, as opposed to recurrent, architecture have been considered \cite{nielsen1997programmable,ghosh2019quantum,ghosh2020universal} and could be used as a starting point.

\section{Methods}

\subsection{Linear network Hamiltonian}\label{sec:Hamiltonian}

We consider a network of interacting quantum harmonic oscillators acting as the reservoir for RC, with spring-like interaction strengths $g_{ij}$. The Hamiltonian of such a system can be conveniently described in terms of the Laplacian matrix $\mathbf{L}$ having elements $\mathbf{L}_{ij}=\delta_{ij}\sum_k g_{ik}-(1-\delta_{ij})g_{ij}$. We adopt such units that the reduced Planck constant $\hbar=1$ and the Boltzmann constant $k_{\mathrm{B}}=1$. Arbitrary units are used for other quantities such as frequency and coupling strength. The resulting Hamiltonian is
\begin{equation}
H=\dfrac{\mathbf{p}^\top\mathbf{p}}{2}+\dfrac{\mathbf{q}^\top(\boldsymbol{\Delta}_{\boldsymbol{\omega}}^2+\mathbf{L})\mathbf{q}}{2},
\label{eq:ham}
\end{equation}
where $\mathbf{p}^\top=\{p_1,p_2,\ldots,p_N\}$ and $\mathbf{q}^\top=\{q_1,q_2,\ldots,q_N\}$ are the vectors of momentum and position operators of the $N$ oscillators while the diagonal matrix $\boldsymbol{\Delta}_{\boldsymbol{\omega}}$ holds the oscillator frequencies $\boldsymbol{\omega}^\top=\{\omega_1,\omega_2,\ldots,\omega_N\}$.

\subsection{Reservoir computing theory}\label{sec:RCtheory}

A common way to process temporal information is to use artificial neural networks with temporal loops. In these so-called recurrent neural networks, the state of the neural network nodes depends on the input temporal signals to be processed but also on the previous states of the network nodes, providing the needed memory \cite{elman1990finding}.
Unfortunately, such recurrent neural networks are notorious for being difficult to train \cite{pascanu2013difficulty}. 
Reservoir Computing, in turn, leads to greatly simplified and faster training, enlarges the set of useful physical systems as reservoirs, and lends itself to simultaneous execution of multiple tasks by training separate output weights for each task while keeping the rest of the network---the reservoir---fixed \cite{lukovsevivcius2009reservoir}.

Here, we provide an overview of Reservoir computing theory that introduces the relevant definitions and concepts in context. For proper development of the discussed material we refer the reader to \cite{konkoli2017reservoir,grigoryeva2018echo}. We will also briefly discuss the application of the framework to quantum reservoirs.

\textbf{Reservoir computers.} We consider sequences of discrete-time data   $\textbf{s}=\{\ldots,\mathbf{s}_{i-1},\mathbf{s}_i,\mathbf{s}_{i+1},\ldots\}$, where $\textbf{s}_{i}\in \mathbb{R}^n$, $n$ is the dimension of the input vector and $i\in\mathbb{Z}$. Let us call the space of input sequences $\mathcal{U}_n$ such that $\mathbf{s}\in \mathcal{U}_n$. Occasionally, we will also use left and right infinite sequences defined as   ${\mathcal{U}_n^-}=\{\mathbf{s}=\{\ldots,\mathbf{s}_{-2},\mathbf{s}_{-1},\mathbf{s}_{0}\}|\mathbf{s}_i\in\mathbb{R}^n,i\in\mathbb{Z}_-\}$ and  ${\mathcal{U}_n^+}=\{\mathbf{s}=\{\mathbf{s}_0,\mathbf{s}_1,\mathbf{s}_2,\ldots\}|\mathbf{s}_i\in\mathbb{R}^n,i\in\mathbb{Z}_+\}$, respectively. Formally, a reservoir computer may be defined by the following set of equations:
\begin{equation}
\begin{cases}
\mathbf{x}_k=T(\mathbf{x}_{k-1},\mathbf{s}_k) \\
o_k=h(\mathbf{x}_k),
\end{cases}
\label{eq:reservoircomputer}
\end{equation}where $T$ is a recurrence relation that transforms input sequence elements $\mathbf{s}_k$ to feature space elements $x_k$---in general, in a way that depends on initial conditions---while $h$ is a function from the feature space to reals. When $T$, a target $\overline{\mathbf{o}}$ and a suitable cost function describing the error between output and target are given, the reservoir is trained by adjusting $h$ to optimize the cost function. The error should remain small also for new input that wasn't used in training. 

The general nature of Eqs.~\eqref{eq:reservoircomputer} makes driven dynamical systems amenable to being used as reservoirs. This has opened the door to so-called physical reservoir computers that are hardware implementations exploiting different physical substrates \cite{tanaka2019recent}. In such a scenario time series $\mathbf{s}$---often after suitable pre-processing---drives the dynamics given by $T$ while $x_k$ is the reservoir state. A readout mechanism that can inspect the reservoir state should be introduced to implement function $h$. The appeal of physical RC lies in the possibility to offload processing of the input in feature space and memory requirements to the reservoir, while keeping the readout mechanism simple and memoryless. In particular, this can lead to efficient computations in terms of speed and energy consumption with photonic or electronic systems \cite{van2017advances,canaday2018rapid}.

\textbf{Temporal maps and tasks.} Online time series processing---what we wish to do with the system in Eqs.~\eqref{eq:reservoircomputer}---is mathematically described as follows. A temporal map {$M:\mathcal{U}_n\to\mathcal{U}_1$} , also called a filter, transforms elements from the space of input time series to the elements of the space of output time series. In general $M$ is taken to be causal, meaning that $(M[\mathbf{s}])_t$ may only depend on $\mathbf{s}_k$ where $k\leq t$, i.e. inputs in the past only. When $M$ is additionally time-invariant, roughly meaning that it does not have an internal clock, $(M[\mathbf{s}])_t=F(\{\ldots,\mathbf{s}_{t-2},\mathbf{s}_{t-1},\mathbf{s}_t\})$ for any $t$ for some fixed {$F:\mathcal{U}^-_n\to\mathbb{R}$} \cite{grigoryeva2018echo}. We will later refer to such $F$ as functionals. When $M$ is given, fixing $\mathbf{s}$ induces a time-dependent function that we will denote by $f$, defined by $f(t)=F(\{\ldots,\mathbf{s}_{t-2},\mathbf{s}_{t-1},\mathbf{s}_t\})$.

To process input $\mathbf{s}$ into $\mathbf{o}$ in an online mode requires to implement $f(t)$; real-time processing is needed. We will later refer to such tasks as temporal tasks. Reservoir computing is particularly suited for this due to the memory of past inputs provided by the recursive nature of $T$ and online processing accomplished by the readout mechanism acting at each time step.

\textbf{Properties of useful reservoirs.} In general, $o_k$ in Eqs.~\eqref{eq:reservoircomputer} depends on both the past inputs and the initial conditions, but $f(t)$ depends only on the inputs; therefore any dependency on the initial conditions should be eliminated by the driving. It may also be expected that reservoirs able to learn temporal tasks must be in some sense well-behaved when driven. These informal notions can be formalized as follows.

The echo state property (ESP) \cite{jaeger2001echo} requires that for any reference time $t$, $\mathbf{x}_t=\mathcal{E}(\{\ldots,\mathbf{s}_{t-2},\mathbf{s}_{t-1},\mathbf{s}_t\})$ for some function $\mathcal{E}$, that is to say at the limit of infinitely many inputs the reservoir state should become completely determined by the inputs, and not by the initial conditions. This has two important consequences. First, it guarantees that the reservoir always eventually converges to the same trajectory of states for a given input, which also means that initial conditions do not need to be taken into account in training. Second, it ensures that the reservoir together with a readout function can realize a temporal map. A strongly related condition called the fading memory property (FMP) \cite{boyd1985fading} requires that for outputs to be similar, it is sufficient that the inputs are similar up to some finite number of past inputs. The formal definition can be given in terms of so-called null sequences as explained in the Supplementary Information. It can be shown that FMP imposes a form of continuity to the overall input-output maps that can be produced by the reservoir computer described by Eqs.~\eqref{eq:reservoircomputer} \cite{maass2004computational}; the associated temporal maps are called fading memory temporal maps.

A useful quantifier for the processing power of a single reservoir was introduced by Dambre et al. \cite{dambre2012information}. They showed that when the readout function $h$ is linear in reservoir variables, the ability of a reservoir to reconstruct orthogonal functions of the input is bounded by the number of linearly independent variables used as arguments of $h$. The central result was that all reservoirs with FMP can saturate this bound.

Considering a class of reservoirs instead offers a complementary point of view. If the reservoirs have ESP and FMP then they can realize temporal maps. If additionally the class has separability, i.e. for any $\mathbf{s}_1,\mathbf{s}_2\in\mathcal{U}^-_n$, $\mathbf{s}_1\neq\mathbf{s}_2$, some reservoir in the class will be driven to different states by these inputs, then universality in RC becomes possible. This can be achieved by imposing mild additional conditions on the input space and realizing an algebra of temporal maps by combining the outputs of multiple reservoirs with a polynomial function \cite{grigoryeva2018universal}. When these properties are met, universality then follows from the Stone-Weierstrass Theorem (Theorem 7.3.1 in Ref. \onlinecite{dieudonne2011foundations}).

\subsection{The explicit forms of the covariance matrix and first moments vector}\label{ap:cov_matrix}

For a single mode Gaussian state with frequency $\Omega$, covariances and first moments read
\begin{equation}
\begin{cases}
\sigma(\mathbf{x})=(n_{th}+\frac{1}{2})\begin{pmatrix}
(y+z_{cos})\Omega^{-1} & z_{sin} \\
z_{sin} & (y-z_{cos})\Omega 
\end{pmatrix},\\
\langle\mathbf{x}\rangle=\begin{pmatrix}
|\alpha|\cos{(\mathrm{arg}(\alpha))}\sqrt{2\Omega^{-1}} \\
|\alpha|\sin{(\mathrm{arg}(\alpha))}\sqrt{2\Omega}
\end{pmatrix},
\end{cases}
\label{eq:methods1}
\end{equation}
where $y=\cosh{(2r)}$, $z_{cos}=\cos{(\varphi)}\sinh{(2r)}$ and $z_{sin}=\sin{(\varphi)}\sinh{(2r)}$. Here, $n_{th}$ controls the amount of thermal excitations, $r$ and $\varphi$ control the magnitude and phase of squeezing, respectively, and finally $|\alpha|$ and $\mathrm{arg}(\alpha)$ control the magnitude and phase of displacement, respectively. The input sequence may be encoded into any of these parameters or possibly their combination.

Suppose that $\mathbf{s}=\{\mathbf{s}_{-m+1},\ldots,\mathbf{s}_{-1},\mathbf{s}_0\}$ and each input $\mathbf{s}_k$ is encoded to all degrees of freedom as $n_{th}\mapsto n_{th}(\mathbf{s}_k)$, $r\mapsto r(\mathbf{s}_k)$, $\varphi\mapsto \varphi(\mathbf{s}_k)$, $|\alpha|\mapsto |\alpha(\mathbf{s}_k)|$ and $\mathrm{arg}(\alpha)\mapsto \mathrm{arg}(\alpha(\mathbf{s}_k))$. Then from Eqs.~\eqref{eq:qrcchannel} it follows that
\begin{widetext}
\begin{equation}
\begin{cases}
\left[\sigma(\mathbf{x}_0^R)-\mathbf{A}^m\sigma(\mathbf{x}_{-m}^R)(\mathbf{A}^\top)^m) \right]_{ij}=\sum_{k=0}^{m-1}a_k^{ij}n_{th}(\mathbf{s}_k)(\cosh{(2r(\mathbf{s}_k))}+(b_k^{ij}\cos{(\varphi(\mathbf{s}_k))}+c_k^{ij}\sin{(\varphi(\mathbf{s}_k))})\sinh{(2r(\mathbf{s}_k))}), \\
\left[\langle\mathbf{x}_0^R\rangle-\mathbf{A}^m\langle\mathbf{x}_{-m}^R\rangle\right]_i=\sum_{k=0}^{m-1}|\alpha(\mathbf{s}_k)|(a_k^{i}\cos{(\mathrm{arg}(\alpha(\mathbf{s}_k)))}+b_k^{i}\sin{(\mathrm{arg}(\alpha(\mathbf{s}_k)))})
\end{cases}
\label{eq:methods2}
\end{equation}
\end{widetext}
where $a_k^{ij}$, $b_k^{ij}$, $c_k^{ij}$, $a_k^{i}$, and $b_k^{i}$ are constants depending on the Hamiltonian in Eq.~\eqref{eq:ham} and $\Delta t$. That is to say the part of the observables independent of the initial conditions $\mathbf{x}_{-m}^R$ are linear combinations of $n_{th}(\mathbf{s}_k)$ and $|\alpha(\mathbf{s}_k)|$, while the dependency on $r(\mathbf{s}_k)$, $\varphi(\mathbf{s}_k)$ and $\mathrm{arg}(\alpha(\mathbf{s}_k))$ is nonlinear. When the dynamics of the reservoir is convergent the effect of the initial conditions vanishes at the limit $m\to\infty$ and the corresponding terms on the L.H.S. may be omitted.

\subsection{The networks used in numerical experiments}\label{sec:usednetworks}
We have used a chain of $N=8$ oscillators for all results shown in Fig.~\ref{fig:predictcomparison} and Fig.~\ref{fig:IPC}. For simplicity, all oscillators have the same frequency $\omega=0.25$ and all interaction strengths are fixed to $g=0.1$. The ancilla is chosen to be one of the oscillators at the ends of the chain. For the aforementioned parameter values of $\omega$ and $g$, we have computed $\rho(\mathbf{A})$ as a function of $\Delta t$. We have set $\Delta t =59.6$, which is close to a local minimum of $\rho(\mathbf{A})$; in general values of $\Delta t$ that achieve $\rho(\mathbf{A})<1$ are common and produce similar results. It should be pointed out that the choice of ancilla matters, e.g., choosing the middle oscillator in a chain of odd length seems to lead to $\rho(\mathbf{A})\geq1$ for any choice of $\Delta t$.

The starting point for the results shown in Fig.~\ref{fig:NLtuning} is a completely connected network of $N=8$ oscillators with uniform frequencies $\omega=0.25$ and random interaction strengths uniformly distributed in the interval $g=[0.01,0.19]$. We point out that the condition $\rho(\mathbf{A})<1$ is favored by interaction strengths that break the symmetry of the system. A suitable value for $\Delta t$ is then found as follows. We consider values $\Delta t \omega_0=0.01,0.02,\ldots,29.99,30$ and find the corresponding $\rho(\mathbf{A})$ for each. Then we choose a random $\Delta t$ out of all values for which $\rho(\mathbf{A})\leq0.99$. In the rare event that none of the values can be chosen, new random weights are drawn and the process is repeated. We have checked that choosing instead the $\Delta t$ that minimizes $\rho(\mathbf{A})$ leads to virtually the same results, confirming that reservoir memory is primarily controlled by the encoding, not the choice of $\Delta t$.

\subsection{Training of the network}\label{sec:training}

For all shown results, we take the initial state of the reservoir to be a thermal state and use the first $10^5$ timesteps to eliminate its effect from the reservoir dynamics, followed by another $M=10^5$ timesteps during which we collect the reservoir observables used to form the output. We wish to find a readout function $h$ that minimizes
\begin{equation}
\mathrm{SE}(\bar{\mathbf{o}},\mathbf{o})=\sum_k(\bar{o}_k-o_k)^2,
\label{eq:SE}
\end{equation}
i.e. the squared error between target output $\bar{\mathbf{o}}$ and actual output $\mathbf{o}$.

In Fig.~\ref{fig:predictcomparison} \textbf{a}, \textbf{b} and in Figs.~\ref{fig:NLtuning} and \ref{fig:IPC}, $h$ is linear in reservoir observables. In this case, the collected data is arranged into $M\times L$ matrix $\mathbf{X}$, where $L$ is the number of used observables. In the case of first moments, $L=2(N-1)$, whereas in the case of covariances $L=2N^2-3N+1$; the latter comes from using the diagonal and upper triangular elements of the symmetric $2(N-1)\times2(N-1)$ covariance matrix. We introduce a constant bias term by extending the matrix with a unit column so that the final dimensions of $\mathbf{X}$ are $M\times (L+1)$. Now we may write $o_k=h(\mathbf{X}_k)=\sum_i^{N+1}W_i X_{ki}$ where $\mathbf{X}_k$ is the $k$th row of $\mathbf{X}$, $X_{ki}$ its $i$th element and $W_i\in\mathbb{R}$ are adjustable weights independent of $k$. Let $\mathbf{W}$ be the column vector of the weights. Now $\mathbf{X}\mathbf{W}=\mathbf{o}^\top$. To minimize \eqref{eq:SE}, we set
\begin{equation}
\mathbf{W}=\mathbf{X}^+\bar{\mathbf{o}}^\top
\end{equation}
where $\mathbf{X}^+$ is the Moore-Penrose inverse \cite{jaeger2002tutorial,lukovsevivcius2012practical} of $\mathbf{X}$. When $\mathbf{X}$ has linearly independent columns---meaning that the reservoir observables are linearly independent---$\mathbf{X}^+=(\mathbf{X}^\top\mathbf{X})^{-1}\mathbf{X}^\top$.

In Fig.~\ref{fig:predictcomparison} \textbf{c}, \textbf{d}, $h$ is taken to be polynomial in reservoir observables. In this case, the training proceeds otherwise as above except that before finding $\mathbf{X}^+$ we expand $\mathbf{X}$ with all possible products of different reservoir observables up to a desired degree, increasing the number of columns. Powers of the same observable are not included since they are not required by the parity check task.

\subsection{The used echo state network}\label{sec:ESN}

An echo state network (ESN) is used for some of the results shown in Fig.~\ref{fig:IPC}. For a comparison with a harmonic network of 8 oscillators (one of them the ancilla) and a bias term we set the size of the ESN to $N=8$ neurons, all of which are used to form the output, and include a bias term.

The ESN has a state vector $\mathbf{x}_k\in\mathbb{R}^N$ with dynamics given by
$\mathbf{x}_k=\tanh (\beta\mathbf{W}\mathbf{x}_{k-1}+\iota\mathbf{w}s_k)$ where $\mathbf{W}$ is a random $N\times N$ matrix, $\mathbf{w}$ a random vector of length $N$, $\beta$ and $\iota$ are scalars and $\tanh$ acts element-wise. $\mathbf{W}$ and $\mathbf{w}$ are created by drawing each of their elements uniformly at random from the interval $[-1,1]$. Furthermore, $\mathbf{W}$ is scaled by dividing it by its largest singular value. Parameters $\beta$ and $\iota$ are used to further adjust the relative importance of the previous state $\mathbf{x}_{k-1}$ and scalar input $s_k$. We use a single fixed realization of $\mathbf{W}$ and $\mathbf{w}$ and set $\beta=0.95$ and $\iota=1$. The readout function is a linear function of the elements of $\mathbf{x}_k$ and training follows a similar procedure to the one described for the oscillator networks in Methods D. 
We note that the precise linear and nonlinear memory contributions of the ESN to the IPC bar in Fig.~\ref{fig:IPC} depend on the choice of the parameters values for $\beta$ and $\iota$. 
For this manuscript, the relevant aspect is that the total IPC of the ESN is bounded to the number of neurons (8), independently of the choice of the parameter values.

\subsection{Estimation of total information processing capacity}\label{sec:IPCmethods}

Information processing capacity is considered in Figs.~\ref{fig:NLtuning} and \ref{fig:IPC}. By total capacity we mean the sum of capacities over a complete orthogonal set of functions and using infinite sequences $\mathbf{s}'$ and $\mathbf{s}$. Shown results are estimates of the total capacity found as follows.

All estimates are formed with input i.i.d. in $[-1,1]$. One choice of functions orthogonal w.r.t. this input is described in Eq.~$(12)$ of Ref. \onlinecite{dambre2012information}, which we also use. More precisely, the considered orthogonality is defined in terms of the scalar product in the Hilbert space of fading memory functions given in Definition $5$ of Ref. \onlinecite{dambre2012information}---it should be stressed that in general, changing the input changes which functions are orthogonal. Since $\sigma(\mathbf{x}^R)$ and $\langle\mathbf{x}^R\rangle$ can only depend on products of the inputs at the same delay, we only consider the corresponding subset of functions. They are of the form $(P_d^\tau)_k=P_d(s_{k-\tau})$ where $P_d$ is the normalized Legendre polynomial of degree $d$ and $\tau\in\mathbb{N}$ is a delay. In Fig.~\ref{fig:IPC}, an estimate for the total capacity of an echo state network is also shown, for which we consider the full set of functions.

For each considered function, we compute the capacity given by Eq.~\eqref{eq:capacity} by finding the weights of the optimal $h$ as described in Methods D. We use finite input sequences, which in general can lead to an overestimation of the total capacity. As explained in the Supplementary Material of Ref. \cite{dambre2012information}, the effect of this can be reduced by fixing a threshold value and setting to $0$ any capacity at or below the value. We use the same method.

In practice, only a finite number of degrees $d$ and delays $\tau$ can be considered for the numerical estimates, which can lead to an underestimation. We have found the following approach useful when searching for capacities larger than the threshold value. We fix a maximum degree $d$ (for all results we have used $9$) and for each degree  we order the functions according to delay and find the capacity of $N/2$ (rounded to an integer) functions at a time, until none of the $N/2$ functions in the batch contribute to total capacity. All total capacities except the one for thermal encoding---where we have verified that some of the observables are in fact linearly dependent or almost---are very close to the theoretical maximum.

A different approach is used for the echo state network, which we briefly describe in the following. We still fix the maximum degree as $9$. For a fixed degree $d$ we consider a sequence of delays $\{\tau_1,\tau_2,\ldots,\tau_d\}$ where the sequence is non-descending to avoid counting the same function multiple times. Then we form the product $\prod_{\tau_i} P_{m(\tau_i)}(s_{k-\tau_i})$ over distinct delays of the sequence where $m(\tau_i)$ is the multiplicity of $\tau_i$ in the sequence. The lexical order of non-descending sequences of delays allows us to order the functions, which is exploited to generate each function just once. Furthermore, we have found that functions that contribute to total capacity seem to have a tendency to be early in the ordering, which makes it faster to get close to saturating the theoretical bound.

\section*{Data availability}

Data is available from the corresponding author at reasonable request.

\bibliography{references}
\bibliographystyle{naturemag}
\section*{Acknowledgements}
We acknowledge the Spanish State Research Agency, through the Severo Ochoa and Mar\'ia de Maeztu Program for Centers and Units of Excellence in R\&D (MDM-2017-0711) and through the  QUARESC project (PID2019-109094GB-C21 and C-22 / AEI / 10.13039/501100011033).
 We also acknowledge funding by CAIB through the QUAREC project (PRD2018/47). The work of MCS has been supported by MICINN/AEI/FEDER and the University of the Balearic Islands through a ``Ramon y Cajal” Fellowship (RYC-2015-18140). VP acknowledges financial support from the European Research Council under the Consolidator Grant COQCOoN (Grant No. 820079).

\section*{Author contributions}
RZ and MCS defined the research topic and supervised the research. JN performed the numerical simulations. RMP, JN and GLG contributed to the theoretical aspects of the paper (demonstrations of Lemma 1 and Universality theorem). VP assessed the potential experimental implementation.
All authors contributed to interpreting the results and writing the article.

\section*{Competing interests}

We declare no competing interests.

\end{document}